\documentclass[twocolumn,prd,amsfonts,showpacs,floats,superscriptaddress,nofootinbib,floatfix,a4paper]{revtex4}

\usepackage[english]{babel}
\usepackage{graphicx}
\usepackage{dcolumn}
\usepackage{bm}
\usepackage{natbib}
\usepackage{amsmath}
\usepackage{amsfonts}
\usepackage{amssymb}
\usepackage{amsthm}

\begin{document}

\title{More about the Tolman-Oppenheimer-Volkoff equations for the generalized Chaplygin gas}
\date{\today}
\author{V. Gorini}
\affiliation{Dipartimento di Fisica e Matematica, Universit\`a dell'Insubria, Via Valleggio 11, 22100 Como, Italy and INFN, sezione di Milano, Via Celoria 16, 20133 Milano, Italy}
\author{A. Yu. Kamenshchik}
\affiliation{Dipartimento di Fisica and INFN, Via Irnerio 46, 40126 Bologna, Italy}
\affiliation{L. D. Landau Institute for Theoretical Physics, Kosygin street 2, 119334 Moscow, Russia}
\author{U. Moschella}
\affiliation{Dipartimento di Fisica e Matematica, Universit\`a dell'Insubria, Via Valleggio 11, 22100 Como, Italy and INFN, sezione di Milano, Via Celoria 16, 20133 Milano, Italy}
\author{O. F. Piattella}
\affiliation{Dipartimento di Fisica e Matematica, Universit\`a dell'Insubria, Via Valleggio 11, 22100 Como, Italy and INFN, sezione di Milano, Via Celoria 16, 20133 Milano, Italy}
\author{A. A. Starobinsky}
\affiliation{L. D. Landau Institute for Theoretical Physics, Kosygin street 2, 119334 Moscow, Russia}

\begin{abstract}
We investigate  the Tolman-Oppenheimer-Volkoff equations for the generalized Chaplygin gas with the aim of extending the findings of V. Gorini, U. Moschella, A. Y. Kamenshchik, V. Pasquier, and A. A. Starobinsky [Phys. Rev. D {\bf 78}, 064064 (2008)]. We study both the standard case, where we reproduce some previous results, and the phantom case. In the phantom case we show that even a superluminal group velocity arising for $\alpha > 1$ cannot prevent the divergence of the pressure at a finite radial distance. Finally, we investigate how a modification of the generalized Chaplygin gas equation of state, required by causality arguments at densities very close to $\Lambda$, affects the results found so far.
\end{abstract}

\maketitle

\section{Introduction}

The presence of dark energy (DE) component in our universe seems to be a matter of fact. In this situation, the study of spherically symmetric solutions of the Einstein equations with additional matter or geometric terms describing DE turns out to be an important issue. Moreover, since the possibility of a phantom DE, namely a dark component which has an equation of state parameter $w < - 1$, is not completely ruled out by observation \cite{Kowalski:2008ez, H09, K09, V09, R09}, this study is even more impelling from the standpoint of wormholelike solutions.

In this paper we generalize the results found in \cite{Gorini:2008zj} to the case of a DE described by the generalized Chaplygin gas (gCg) \cite{Kamenshchik:2001cp} (see also \cite{BBS02}), in particular, in the light of our recent paper \cite{Gorini:2007ta}, where the possibility of a {\it superluminal} regime has been noticed; in \cite{Gorini:2007ta} we have shown that the superluminal regime does not contradict causality provided some modifications of the gCg equation of state are made which do not affect the present and past evolution of the universe. It is also interesting to investigate the effects of a very large gCg group velocity on the geometries already found in \cite{Gorini:2008zj}.

In Sec. \ref{sec:toveqs}, we write down the Tolman-Oppenheimer-Volkoff (TOV) equations, and in Sec. \ref{sec:constprsol} we exhibit their special constant pressure solutions. Sections \ref{sec:Nonphant} and \ref{sec:phant} are devoted to the analysis of the normal case $|p| < \rho$ and of the phantom one $|p| > \rho$, respectively. In Sec. \ref{sec:superluminalityissue} we address the superluminality issue and we investigate how a modification of the gCg equation of state changes the pressure solution without spoiling the results found. Section \ref{sec:concl} is devoted to discussion and conclusions; in particular, we emphasize the differences arising in the gCg model compared to the standard Cg ($\alpha = 1$) case.
\newpage

\section{Tolman-Oppenheimer-Volkoff equations in the presence of the generalized Chaplygin gas}\label{sec:toveqs}

We assume a static spherically symmetric space-time geometry ($c = 8\pi G = 1$):\footnote{In the preceding paper \cite{Gorini:2008zj} we have used a different normalization: $G = 1$.}
\begin{equation}
ds^{2} = e^{\nu(r)}dt^{2} - e^{\mu(r)}dr^{2} - r^{2}\left(d\theta^{2} + \sin^{2}\theta d\phi^{2}\right)
\end{equation}
and a perfect fluid stress-energy tensor
\begin{equation}
T_{\mu\nu} = (\rho + p)u_{\mu}u_{\nu} - g_{\mu\nu}p\;.
\end{equation}
Einstein's equations, conservation of the energy and the boundary condition
\begin{equation}
e^{-\mu(0)} = 1\;
\end{equation}
together imply the TOV \cite{Tolman:1939jz, Oppenheimer:1939ne} differential equation
\begin{equation}
\frac{dp}{dr} = - \frac{(\rho + p)(M + 4\pi r^{3}p)}{r(8\pi r - 2M)}\;,
\end{equation}
where
\begin{equation}
 M(r) = \int_{0}^{r} \rho(u)\; 4\pi{u}^{2}\; du
\end{equation}
satisfies the equation
\begin{equation}
 \frac{dM}{dr} = 4\pi r^{2}\rho\;, \  \  \ M(0) = 0\;.
\end{equation}
The equation of state of the gCg \cite{Kamenshchik:2001cp, BBS02} is given by
\begin{equation}\label{eos}
p = -\frac{\Lambda^{\alpha + 1}}{\rho^{\alpha}}\;,
\end{equation}
where $\Lambda$ and $\alpha$ are positive constants. By plugging (\ref{eos}) in the TOV equation we get the following first-order system of differential equations:
\begin{eqnarray}
\label{presseq}\frac{d|p|}{dr} & = & \frac{\left(\Lambda^{\beta + 1} - |p|^{\beta + 1}\right)\left(M - 4\pi r^{3}|p|\right)}{|p|^{\beta}r(8\pi r - 2M)}\;,\\
\label{masseq}
\frac{dM}{dr} & = & 4\pi r^{2}\frac{\Lambda^{\beta + 1}}{|p|^{\beta}}
\end{eqnarray}
written in terms of the modulus of the pressure and of the parameter $\beta \equiv 1/\alpha$.

\section{Constant pressure solutions}\label{sec:constprsol}

One solution with constant pressure is the de Sitter universe:
\begin{equation}
\begin{array}{ll}
p = - \rho = - \Lambda\;, & M(r) = \frac{4}{3}\pi\Lambda r^{3}\;,\\
\end{array}
\end{equation}
where the de Sitter radius is given by
\begin{equation}
r_{dS} = \sqrt{\frac{3}{\Lambda}}
\end{equation}
and the curvature does not depend on $\beta$.

The Einstein static universe (ESU) is another solution:
\begin{equation}
\begin{array}{rll}
p &=& -\Lambda_{\rm eff} \equiv - {3}^{-\frac{1}{\beta + 1}}\Lambda\;,
\\ \rho &=& \Lambda_{\rm eff} + \bar \rho =  3^{\frac{\beta}{\beta + 1}}\Lambda\;, \ \ \ \ \bar \rho = 2 \Lambda_{\rm eff} \;,\\ & & \\ M(r) &=& 4 \pi \Lambda_{\rm eff}\, r^{3}\;.
\end{array}
\end{equation}
The radius of the spatial spherical section of the ESU is
\begin{equation}
r_{E} = \sqrt{\frac{3^{\frac{\beta}{\beta + 1}}}{\Lambda}}\;. \label{pp}
\end{equation}
In the limit $\beta \to \infty$ the two radii  $r_{E}$ and $ r_{dS}$  coincide (but not the geometries), while in the ultrasuperluminal $\beta \to 0$ limit $r_{E} \to \sqrt{{1}/{\Lambda}}\;$.

\section{The non-phantom case $\rho > |p|$}\label{sec:Nonphant}

Consider the system (\ref{presseq})-(\ref{masseq}) and assume the dominant energy condition $\rho > |p|$.

We fix the pressure $p(r_b)$ and the mass $M(r_b)$ at some value $r = r_b$. The following inequality holds: \begin{equation}
4\pi  r_b > M(r_b)\;. \label{condA}
\end{equation}
At these initial conditions the pressure can neither vanish nor attain the value $p = -\Lambda$. The former case is excluded because, when $p \to 0$, the right-hand side (rhs) of Eq. (\ref{presseq}) is positive and therefore the pressure cannot vanish. The value $p = -\Lambda$ is also excluded, which can be seen by rewriting Eq. (\ref{presseq}) as follows:
\begin{equation}
d\ln\left(|p|^{\beta + 1} - \Lambda^{\beta + 1}\right) = -(\beta + 1)\frac{M - 4\pi r^{3}|p|}{r(8\pi r - 2M)}dr\;.
\end{equation}
If the pressure could approach $-\Lambda$, the left-hand side (lhs) would diverge logarithmically while the rhs would stay regular: a contradiction.

If the condition $\rho > |p|$ is satisfied, then $\rho(r) > \Lambda$ or equivalently
\begin{equation}
-\Lambda < p(r) < 0 \label{condB}\;.
\end{equation}
Equation (\ref{masseq}) then implies that $M(r)$ grows at least as fast as $r^{3}$ and, therefore, the equality $4 \pi r_{0} = M(r_{0})$ is achieved at a certain radius $r_{0}$.

At $r = r_0$ the denominator on the rhs of Eq. (\ref{presseq}) diverges unless ${r_{0}^{2}}p(r_{0}) = -{1}$.

To elaborate on this point let us rewrite Eq. (\ref{presseq}) by expanding the relevant quantities in the neighborhood of $r_{0}$. Let $p_0$ and $\rho_0$ be the values of the pressure and of the energy density at $r_0$. To first order the mass is given by
\begin{equation}
M(r) = 4\pi {r_{0}} - 4\pi r_{0}^{2}\frac{\Lambda^{\beta + 1}}{|p_{0}|^{\beta}}\, \epsilon\;,
\end{equation}
where we have set $\epsilon = r_{0} - r$. As for the pressure, we have
\begin{equation}
|p(r)| = |p_{0}| + \tilde{p}(\epsilon)\;,
\end{equation}
where $\tilde{p}(\epsilon)$ vanishes when $\epsilon \to 0$.

Hence, for $\epsilon \simeq 0$ Eq. (\ref{presseq}) takes the following approximate form:
\begin{equation}
\frac{d\tilde{p}}{d\epsilon} = \frac{\left(\Lambda^{\beta + 1} - |p_{0}|^{\beta + 1}\right)\left(1 -  r_{0}^{2}|p_{0}|\right)}{2\epsilon\left(|p_{0}|^{\beta} - r_{0}^{2}\Lambda^{\beta + 1}\right)}\;.
\end{equation}
A logarithmic divergence is manifest unless
\begin{equation}
|p_{0}| = \frac{1}{r_{0}^{2}}\;.
\end{equation}
The inequality $-\Lambda < p(r) < 0$ provides a lower bound for $r_{0}$:
\begin{equation}
-\Lambda < p_{0} = -\frac{1}{r_{0}^{2}} \ \rightarrow \ r_{0} > \sqrt{\frac{1}{\Lambda}}\;;
\end{equation}
the inequality  $\rho(r) > \Lambda$ implies a bound for the mass function:
\begin{equation}
M(r) > \frac{4}{3}\pi r^{3}\Lambda\;.
\end{equation}
Since $M(r_{0}) = 4\pi r_{0}$ it follows that
\begin{equation}
\sqrt{\frac{1}{\Lambda}} < r_{0} < \sqrt{\frac{3}{\Lambda}} \label{rrr}
\end{equation}
[see Eq. (\ref{pp})]. The approximate equation for $\tilde{p}$ is written
\begin{equation}\label{eqepsilon}
\frac{d\tilde{p}}{d\epsilon} = \frac{\tilde{p}}{2\epsilon} + C_{0}\;,
\end{equation}
where
\begin{equation}
C_{0} = \frac{1}{2r_{0}^{3}}\left[\left(r_{0}^{2}\Lambda\right)^{\beta + 1} - 3\right]\;,
\end{equation}
so that
\begin{equation}
\tilde{p} = A\sqrt{\epsilon} + 2C_{0}\epsilon\;,
\end{equation}
where $A$ is an integration constant. The solutions $p(r), M(r)$ are characterized by the parameters $A$ and $r_{0}$ which are in turn determined by the boundary conditions at  $r = r_{b}$. 

In order to avoid the apparent singularity on the rhs of Eq. (\ref{presseq}) it is useful to change the radial coordinate as follows:
\begin{equation}
r = r_{0}\sin\chi\;.
\end{equation}
The gCg TOV equations take the following form:
\begin{eqnarray}
\frac{d|p|}{d\chi} & = & \frac{\left(\Lambda^{\beta + 1} - |p|^{\beta + 1}\right)\left(M - 4\pi r_{0}^{3}|p|\sin^{3}\chi\right)\cos\chi}{|p|^{\beta}\sin\chi(8\pi r_{0}\sin\chi - 2M)}~\;,\cr
& & \label{presseqchi} \\
\label{Masseqchi}
\frac{dM}{d\chi} & = & 4\pi r_{0}^{3}\sin^{2}\chi\cos\chi\frac{\Lambda^{\beta + 1}}{|p|^{\beta}}\;.
\end{eqnarray}
We investigate these equations in the vicinity of the equator ($\chi = \frac{\pi}{2}$) by the substitution
\begin{equation}
\chi = \frac{\pi}{2} - \delta\;,
\end{equation}
where $\delta$ is positive. Once more, we expand the pressure and the mass,
\begin{equation}
|p| = \frac{1}{r_{0}^{2}} + \bar{p}(\delta)\;,
\end{equation}
\begin{equation}
M = 4\pi r_{0} - 2\pi r_{0}\left(r_{0}^{2}\Lambda\right)^{\beta + 1}\delta^{2}
\end{equation}
to lowest order in $\delta$. We can then write an approximate equation for $\bar{p}(\delta)$:
\begin{equation}\label{eqalpha}
\frac{d\bar{p}}{d\delta} = \frac{\bar{p}}{\delta} + C_{1}\delta\;,
\end{equation}
where
\begin{equation}
C_{1} = \frac{1}{2 r_{0}^{2}}\left[\left(r_{0}^{2}\Lambda\right)^{\beta + 1} - 3\right]\;.
\end{equation}
The solution of (\ref{eqalpha}) is
\begin{equation}\label{solalpha}
\bar{p} = B\delta + C_{1}\delta^{2}\;,
\end{equation}
where $B$ is an integration constant. The presence of the parameter $\beta$ does not appreciably affect the equator crossing being of subleading order in $\delta$. The Einstein static solution corresponds to the choice $B = 0$ and $C_{1} = 0$:
\begin{equation}
r_{0}^{2} = \frac{3^{\frac{1}{\beta + 1}}}{\Lambda}\;.
\end{equation}
The solution (\ref{solalpha}) can now be continued to negative values of $\delta$ and can describe the equator crossing. This can be done by introducing the variable
\begin{equation}
y = \frac{1}{\sin\chi}\;,
\end{equation}
so that $y \in [1,\infty)$. Equations (\ref{presseqchi})-(\ref{Masseqchi}) become
\begin{eqnarray}
\frac{d|p|}{dy} & = & - \frac{\left(\Lambda^{\beta + 1} - |p|^{\beta + 1}\right)\left(My^{3} - 4\pi r_{0}^{3}|p|\right)}{|p|^{\beta}y^{3}(8\pi r_{0} - 2My)}\;,
\label{press-eq}\\
\frac{dM}{dy} & = &  - \frac{4\pi r_{0}^{3}\Lambda^{\beta + 1}}{y^{4}|p|^{\beta}}\;.
\label{mass-eq}
\end{eqnarray}
Following \cite{Gorini:2008zj} we now show that the quantity $8\pi r_{0} - 2My$  is always positive.

The rhs of Eq. (\ref{mass-eq}) attains the smallest possible value at $p = -\Lambda$; this implies that
\begin{equation}
8\pi r_{0} - 2My \geqslant 8\pi r_{0} - \frac{8\pi\Lambda r_{0}^{3}}{3y^{2}}\;.
\end{equation}

The rhs is positive for
\begin{equation}
y^{2} > \frac{\Lambda r_{0}^{2}}{3}\;. \label{condC}
\end{equation}
Since by definition  $y^{2} > 1$, the inequalities (\ref{rrr}) and (\ref{condC}) imply that $8\pi r_{0} - 2My$ is always positive.

We are now in a position to investigate solutions of the TOV equations for $\chi \to \pi$, i.e. $y \to \infty$.

\paragraph*{Case 1.} Consider solutions for which the pressure attains, at infinity, a generic value in the range $0 < |p_{\infty}| < \Lambda$. For such a solution to exist, the integral
\begin{equation}
\int^{y}_{y_b} dv\frac{Mv^{3} - 4\pi r_{0}^{3}|p|}{v^{3}(8\pi r_{0} - 2Mv)}\;,
\end{equation}
must converge for $y \to \infty$, where $y_b = r_0/r_b$; $M_{\infty}$ cannot be either positive (otherwise $8\pi r_{0} - 2Mv$ would become negative) or negative (otherwise the integral would diverge logarithmically). Therefore, $M_{\infty}$ must necessarily vanish. Then, Eq. (\ref{mass-eq}) gives the asymptotic behavior
\begin{equation}
M \simeq \frac{4\pi r_{0}^{3}\Lambda^{\beta + 1}}{3y^{3}|p_{\infty}|^{\beta}}\;,
\end{equation}
and, from Eq. (\ref{press-eq}),
\begin{equation}
\frac{d|p|}{dy} \simeq -r_{0}^{2}\frac{\left(\Lambda^{\beta + 1} - |p_{\infty}|^{\beta + 1}\right)\left(\Lambda^{\beta + 1} - 3|p_{\infty}|^{\beta + 1}\right)}{6|p_{\infty}|^{2\beta}y^{3}}\;.
\end{equation}

\paragraph*{Case 2.} $|p(y_a)| = 0$ at a certain $y_a > 1$. Let $M(y_a) = M_a$. In the neighborhood of $y_a$ Eq. (\ref{press-eq}) can be written as
\begin{equation}
\frac{d|p|}{dy} \simeq -\frac{\Lambda^{\beta + 1}M_a}{|p|^{\beta}(8\pi r_{0} - 2M_ay_a)}\;. \label{opop}
\end{equation}
In order to have a nonpositive derivative it is necessary that $M_a \geqslant 0$. First consider the case $M_a \neq 0$. Assume the following asymptotic behavior for the pressure in the neighborhood of $y_a$:
\begin{equation}
|p| \simeq D(y_a - y)^{\gamma}\;,
\end{equation}
where $D$ and $\gamma$ are positive constants. Equation (\ref{opop}) gives
\begin{equation}
D\gamma(y_a - y)^{\gamma - 1} = \frac{\Lambda^{\beta + 1}M_a}{D^{\beta}(y_a - y)^{\gamma\beta}
(8\pi r_{0} - 2M_ay_a)}
\end{equation}
and therefore
\begin{equation}
\gamma = \frac{1}{\beta + 1}\;,
\end{equation}
\begin{equation}
D^{\beta + 1} = \frac{(\beta + 1)\Lambda^{\beta + 1}M_a}{8\pi r_{0} - 2M_ay_a}\;.
\end{equation}
If $M_a = 0$ we assume near $y_a$:
\begin{equation}
M(y) \simeq M_{1}(y_a - y)^{\zeta}\;,
\label{zeta}
\end{equation}
\begin{equation}
|p| \simeq E(y_a - y)^{\sigma}\;,
\label{sigma}
\end{equation}
where $M_{1}$ and $E$ are positive constants and $0 < \zeta < \sigma$.\footnote{This inequality comes from Eqs. (\ref{press-eq})-(\ref{mass-eq}) written as
$$
\frac{d|p|}{dM} = \frac{\left(\Lambda^{\beta + 1} - |p|^{\beta + 1}\right)\left(My^{3} - 4\pi r_{0}^{3}|p|\right)y}{4\pi r_{0}^{3}\Lambda^{\beta + 1}(8\pi r_{0} - 2My)}
$$
which vanishes for $y \to y_a$.}

Substituting (\ref{zeta})-(\ref{sigma}) into Eq. (\ref{mass-eq}) we get
\begin{equation}
M_{1}\zeta(y_a - y)^{\zeta - 1} = \frac{4\pi r_{0}^{3}\Lambda^{\beta + 1}}{y_a^{4}E^{\beta}(y_a - y)^{\beta\sigma}}\;.
\end{equation}
It follows that
\begin{equation}\label{mebeta}
M_{1}E^{\beta} = \frac{4\pi r_{0}^{3}\Lambda^{\beta + 1}}{y_a^{4}\zeta}
\end{equation}
and
\begin{equation}\label{zetabetasigma}
\zeta = 1 - \beta\sigma\;.
\end{equation}
Since $\zeta > 0$ we can infer that
\begin{equation}
\frac{1}{\beta + 1} < \sigma < \frac{1}{\beta}\;.
\label{betasigma}
\end{equation}
Plugging (\ref{zeta}) and (\ref{sigma}) into Eq. (\ref{opop}) one finds
\begin{equation}
E\sigma(y_a - y)^{\sigma - 1} \simeq \frac{\Lambda^{\beta + 1}M_{1}}{E^{\beta}(y_a - y)^{\sigma\beta - \zeta}8\pi r_{0}}\;.
\end{equation}
Therefore, it follows that
\begin{equation}\label{sigmazetabeta}
\sigma = \frac{1 + \zeta}{\beta + 1}
\end{equation}
and
\begin{equation}\label{ebeta+1}
E^{\beta + 1} = \frac{\Lambda^{\beta + 1}M_{1}}{8\pi r_{0}\sigma}\;.
\end{equation}
Combining (\ref{sigmazetabeta})-(\ref{ebeta+1}) with (\ref{mebeta})-(\ref{zetabetasigma}) we obtain
\begin{eqnarray}
\sigma &=& \frac{2}{1 + 2\beta}\;, \nonumber\\
\zeta &=& \frac{1}{1 + 2\beta}\;,
\end{eqnarray}
and
\begin{eqnarray}
M_{1} &=& \frac{16\pi r_{0}}{\Lambda^{\beta + 1}(1 + 2\beta)}E^{\beta + 1}, \nonumber\\
E^{2\beta + 1} &=& \frac{r_{0}^{2}\Lambda^{2(\beta + 1)}(1 + 2\beta)^{2}}{4y_a^{4}}\;.
\end{eqnarray}

\paragraph*{Case 3.} $p_{\infty} = -\Lambda$. In this case, rewrite the pressure as follows:
\begin{equation}
|p| = \Lambda - |\tilde{p}|\;.
\end{equation}
Equation (\ref{press-eq}) has the following asymptotic form:
\begin{equation}
\frac{d|\tilde{p}|}{dy} \simeq -\frac{(\beta + 1)|\tilde{p}|}{2y}\;,
\end{equation}
whose solution is
\begin{equation}
|\tilde{p}| = \frac{F}{y^{\frac{\beta + 1}{2}}}\;,
\end{equation}
where $F$ is a positive integration constant.

\section{The phantom case $\rho < |p|$}\label{sec:phant}

The violation of the dominant energy condition is interesting for the following two reasons:
\begin{enumerate}
\item $\rho < |p|$ is a necessary condition for the existence of wormholes solutions \cite{Morris:1988cz, Morris:1988tu} (but it is not sufficient for their stability \cite{Bronnikov:2006pt}).
\item The possibility of a phantom DE has not been ruled out by observation \cite{Kowalski:2008ez,H09,K09,V09,R09}.
\end{enumerate}
We consider Eqs. (\ref{presseq})-(\ref{masseq}) together with the assumption
\begin{equation}\label{phantcond}
p(r_{b}) < -\Lambda\;.
\end{equation}

\paragraph*{Case A.}

If
\begin{equation}\label{caseA}
M(r_{b}) - 4\pi r_{b}^{3}|p(r_{b})| < 0
\end{equation}
then $d|p|/dr > 0$ at $r = r_{b}$. Since $|p|$ grows and satisfies condition (\ref{phantcond}), the term $M - 4\pi r^{3}|p|$ stays negative also for $r > r_{b}$. Therefore, $d|p|/dr > 0$ for $r > r_{b}$. Then, there are three possible asymptotic behaviors:
\begin{enumerate}
\item For $r \to \infty$, $|p|$ tends to a certain finite value $|p(\infty)| > \Lambda$.
\item For $r \to \infty$, $|p|$ diverges.
\item For $r$ tending to a finite value, say $r_{1}$, $|p|$ diverges.
\end{enumerate}
The first subcase is ruled out since the rhs of Eq. (\ref{presseq}) diverges for $r \to \infty$.

The second subcase cannot take place because Eq. (\ref{presseq}) has the following asymptotic form:
\begin{equation}
\frac{d|p|}{dr} \simeq \frac{1}{2} r|p|^{2}\;,
\end{equation}
whose solution
\begin{equation}
|p| \simeq \frac{4|p(r_{b})|}{4 + |p(r_{b})|\left(r_b^{2} - r^2\right)}
\end{equation}
diverges at the finite distance $r^2 = r_b^2 + 4/|p(r_{b})|$. This is a contradiction.

Then, we are left with subcase 3. In the neighborhood of $r = r_1$, assume for the pressure the following power law behavior:
\begin{equation}
|p| = \frac{p_{1}}{(r_{1} - r)^{\eta}}\;,
\end{equation}
where $\eta > 0$. Let $M(r_1) = M_1$; Eq. (\ref{presseq}) has the following form in the neighborhood of $r = r_1$:

\begin{equation}
\frac{\eta p_{1}}{(r_{1} - r)^{\eta + 1}} = \frac{4\pi r_{1}^{2}p_{1}^{2}}{8\pi r_{1} - 2M_{1}}(r_{1} - r)^{-2\eta}\;,
\end{equation}
from which
\begin{equation}\label{etaparam}
\eta = 1
\end{equation}
and
\begin{equation}\label{p1param}
p_{1} = \frac{8\pi r_{1} - 2M_{1}}{4\pi r_{1}^{2}}\;.
\end{equation}

\paragraph*{Case B.}

If
\begin{equation}\label{caseB}
M(r_{b}) - 4\pi r_{b}^{3}|p(r_{b})| > 0\;,
\end{equation}
then $d|p|/dr < 0$ at $r = r_{b}$. From Eq. (\ref{presseq}) it is easily seen that $|p|$ cannot decrease {\it ad libitum} because at a certain finite radius, say $r_2$, the term $M - 4\pi r^{3}|p|$ changes sign, becoming negative.

Then, the pressure has the following profile: from $p(r_b)$ it grows up to a certain (negative) value, say $p_{\rm max}$ at $r = r_{2}$, and then decreases according to the behavior described in subcase 3 of A, diverging at $r = r_1$ with $r_{1} > r_{2}$.

In general, $p_{\rm max} \leqslant -\Lambda$. We prove that the equality is ruled out. We rewrite the pressure equation as follows:

\begin{equation}\label{approxpeq}
d\ln\left(|p|^{\beta + 1} - \Lambda^{\beta + 1}\right) = -(\beta + 1)\frac{M - 4\pi r^{3}|p|}{r(8\pi r - 2M)}dr\;.
\end{equation}
If $M - 4\pi r^{3}|p| > 0$ and $8\pi r - 2M > 0$, we have a logarithmic divergence on the lhs while the rhs is regular, i.e. a contradiction.

If we demand that $8\pi r - 2M$ vanishes in $r_{2}$, Eq. (\ref{approxpeq}) has the following form in the neighborhood of $r = r_2$:
\begin{equation}
d\ln\left(|p|^{\beta + 1} - \Lambda^{\beta + 1}\right) = -\frac{\beta + 1}{2}\frac{dr}{r - r_{2}}\;,
\end{equation}
whose solution is
\begin{equation}
|p|^{\beta + 1} - \Lambda^{\beta + 1} \simeq \left|r - r_2\right|^{-\frac{\beta + 1}{2}}\;.
\end{equation}
For $r \to r_2$ the lhs vanishes while the rhs diverges, i.e. another contradiction.

The last possibility is that $M - 4\pi r^{3}|p| = 0$ and $8\pi r - 2M = 0$ at $r = r_2$. In this case
\begin{equation}
p = -\frac{1}{r_{2}^{2}}\;
\end{equation}
and from Eq. (\ref{masseq})
\begin{equation}
8\pi r - 2M \simeq 8\pi(r - r_{2})\left[1 - \left(r_{2}^{2}\Lambda\right)^{\beta + 1}\right]\;.
\end{equation}
Since $|p| > \Lambda$ and $r_{2}^{2}\Lambda < 1$ the rhs is negative, while $8\pi r - 2M > 0$ by assumption, which is again a contradiction.

In conclusion, only two regimes are possible for a stellar object immersed in the phantom gCg:
\begin{enumerate}
\item If the boundary conditions at the surface of this object satisfy Eq. (\ref{caseA}), then $|p|$ grows and diverges at a finite radius $r = r_{1}$.

\item If instead inequality (\ref{caseB}) holds true, then the pressure grows, attaining a maximum (negative) value $p_{\rm max} < -\Lambda$ at $r = r_2$, and then it decreases, diverging at $r = r_1$.
\end{enumerate}

It is important to stress that, as it can be seen from Eqs. (\ref{etaparam})-(\ref{p1param}), the divergence of the pressure at a finite radius does not depend on the parameter $\beta$. Therefore, the formation of a curvature singularity at a finite value of $r$ is unavoidable even for a very large gCg sound speed.

Then we are in the position to generalize the theorem in \cite{Gorini:2008zj}:

\paragraph*{In a static spherically symmetric universe filled with the phantom generalized Chaplygin gas, the scalar curvature becomes singular at some finite value of the radial coordinate and the universe is not asymptotically flat.}

\subsection{Wormhole-like solutions for a universe filled exclusively with the phantom generalized Chaplygin gas}

Let us consider $8\pi r_{b} - 2M_{b} > 0$ and $p_{b} < -\Lambda$ and investigate the solution of Eq. (\ref{presseq}) for small values of $r$. There are two possibilities:

\begin{enumerate}
\item $8\pi r - 2M(r)$ remains positive up to $r = 0$ ;

\item $8\pi r - 2M(r)$ vanishes at a certain $r = r_{0}$ .
\end{enumerate}
In the first case, let $p(r = 0) = p_0$ and, using Eqs. (\ref{presseq})-(\ref{masseq}), we expand the pressure modulus and the mass in the neighborhood of $r = 0$:
\begin{eqnarray}
|p| &=& |p_{0}|\nonumber\\
&+& \frac{\left(|p_{0}|^{\beta + 1} - \Lambda^{\beta + 1}\right)\left(3|p_{0}|^{\beta + 1} - \Lambda^{\beta + 1}\right) r^{2}}{12|p_{0}|^{2\beta}}\;, \\
M &=& \frac{4\pi \Lambda^{\beta + 1}}{3|p_{0}|^{\beta}}r^{3}\;.
\end{eqnarray}
In the second case $8\pi r - 2M$ vanishes at $r = r_{0}$. As for the nonphantom case of Sec. \ref{sec:Nonphant}, $M - 4\pi r^{3}|p|$ vanishes at $r_{0}$ and $p_{0}r_{0}^{2} = -1$. While in the nonphantom case $r_{0}$ is the maximum value of the radial coordinate, in the phantom case it is the minimum one and represents the radius of a throat.

We now investigate the crossing of the latter. To this purpose, we change coordinates as follows:
\begin{equation}
r = r_{0}\cosh\eta\;.
\end{equation}
Eqs. (\ref{presseq})-(\ref{masseq}) become
\begin{eqnarray}\label{presseqeta}
\frac{d|p|}{d\eta} & = & \frac{\left(\Lambda^{\beta + 1} - |p|^{\beta + 1}\right)\left(M - 4\pi r_{0}^{3}|p| \cosh^{3}\eta\right)\sinh\eta}{|p|^{\beta}\cosh\eta(8\pi r_{0}\cosh\eta - 2M)}\;,\nonumber\\
\frac{dM}{d\eta} & = & 4\pi r_{0}^{3}\cosh^{2}\eta\sinh\eta\frac{\Lambda^{\beta + 1}}{|p|^{\beta}}\;.
\end{eqnarray}
The solution of the mass equation near $r_{0}$, i.e. for small $\eta$, is
\begin{equation}
M = 4\pi r_{0}\left[1 + \frac{1}{2}\left(r_{0}^{2}\Lambda\right)^{\beta + 1}\eta^{2}\right]\;,
\end{equation}
while, rewriting the pressure as
\begin{equation}
p = -\frac{1}{r_{0}^{2}} + \tilde{p}\;,
\end{equation}
Eq. (\ref{presseqeta}) takes the following form near the throat:
\begin{equation}
\frac{d\tilde{p}}{d\eta} = \frac{\tilde{p}}{\eta} + C_{T}\eta\;,
\end{equation}
where
\begin{equation}\label{ctconst}
C_{T} = \frac{3 - \left(r_{0}^{2}\Lambda\right)^{\beta + 1}}{2 r_{0}^{2}}\;.
\end{equation}
The solution of this equation is
\begin{equation}
\tilde{p} = D\eta + C_{T}\eta^{2}\;,
\end{equation}
where $D$ is an arbitrary constant. It is important to emphasize that the wormholelike solutions that we have found here are completely different from the Morris-Thorne-Yurtsever ones \cite{Morris:1988tu} since they do not connect two asymptotically flat space-time regions, but two regions in which the space-time has a singularity.

\section{The superluminal non-phantom generalized Chaplygin gas and the causality problem}\label{sec:superluminalityissue}

The gCg speed of sound is
\begin{equation}\label{cs2gcgeq}
c_{s}^{2} = \frac{\alpha\Lambda^{\alpha + 1}}{\rho^{\alpha + 1}}\;.
\end{equation}
In the range $\alpha > 1$ (i.e. $\beta < 1$) it can exceed the speed of light for values of the energy density sufficiently close to $\Lambda$. This occurs for
\begin{equation}\label{condcaus}
\rho < \alpha^{\frac{1}{\alpha + 1}}\Lambda \equiv \rho_{\rm sl}\;.
\end{equation}
The largest value of $\rho_{\rm sl}$ is attained at the maximum of the function $\alpha^{\frac{1}{\alpha + 1}}$, namely at
\begin{equation}
\alpha_{\rm max} = \exp\left[\mbox{W}\left(\frac{1}{e}\right) + 1\right] \approx 3.591\;,
\end{equation}
where W is the Lambert function. At $\alpha = \alpha_{\rm max}$ one has $\rho_{sl} \approx 1.321\Lambda$.

In cosmology a speed of sound exceeding 1 does not contradict causality. Indeed, what is required by causality is that the signal (i.e. the wavefront) velocity does not exceed that of light. As we have shown in \cite{Gorini:2007ta} using an underlying tachyonlike field model for the gCg, the condition that $c_s\le 1$ for the asymptotic "vacuum" state $\rho = -p = \Lambda$ is sufficient for the purpose. Therefore, following \cite{Gorini:2007ta}, we solve the causality problem for $\alpha > 1$ by smoothly modifying the gCg equation of state to that of the standard Chaplygin gas ($\alpha = 1$) at some value $\rho = \rho_c$ very close to $\Lambda$.

If we assume that $\rho_c < \rho_0$, where $\rho_0$ is the present cosmological energy density, the required change in the equation of state will occur in our future. However, this {\it escamotage} does not work in the geometries described in the present paper because the change takes place at a finite coordinate distance $r$.

In order to quantitatively describe the transition in the equation of state, assume that $\rho = \rho_c$ for $r = r_{c}$ and suppose that $\alpha$ is now a function of the energy density so that $\alpha \to 1$ sufficiently fast for $\Lambda < \rho < \rho_{c}$ and it becomes constant for $\rho > \rho_{c}$. A simple example of such function is
\begin{equation}
\alpha = 1 + \frac{\bar{\alpha}}{2}\left[\tanh\left(\frac{\rho - \rho_{c}}{\delta}\right) + 1\right]\;,
\end{equation}
where $\delta$ is a small (i.e. $|\Lambda - \rho_c| \gg \delta$) positive parameter which tunes the rapidity of the transition and $\bar{\alpha}$ is a constant.

We now investigate the effect of the transition in the equation of state on the pressure and mass configurations in order to understand if all the results found so far still hold true.

Assume that $\rho_{c}$ is close to $\Lambda$, i.e.
\begin{equation}\label{xdef}
x \equiv \frac{\rho_{c} - \Lambda}{\Lambda} \ll 1\;.
\end{equation}
Then from (\ref{eos}) we have
\begin{equation}\label{pc}
|p_{c}| = \Lambda\left[1 - \alpha\left(r_{c}\right) x\right]\;,
\end{equation}
to first order in $x$.
Equation (\ref{presseq}) now reads
\begin{equation}\label{presseqrhoc}
\frac{d|p|}{dr} = \frac{\left(\Lambda^{\beta(r) + 1} - |p|^{\beta(r) + 1}\right)\left(M - 4\pi r^{3}|p|\right)}{|p|^{\beta(r)}r(8\pi r - 2M)}
\end{equation}
with a $r$-dependent parameter $\beta$. We expand all the relevant quantities in the neighborhood of $r_{c}$:
\begin{equation}\label{expansions}
\begin{array}{lll}
 r  &=& r_{c} + \epsilon\;,\\ \\
 M(r) &=& M_{c} + 4\pi r_{c}^{2}\rho_{c}\epsilon\;,\\ \\
|p|(r) &=& |p_{c}| + \tilde{p}(\epsilon)\;,\\ \\
\beta(r) &=& \beta_{c} + \left.\frac{\partial\beta}{\partial r}\right|_{r_{c}}\epsilon\;.
\end{array}
\end{equation}
Assume that $\beta_{c} = 1$ and substitute formulas (\ref{xdef})-(\ref{pc}) and the expansions (\ref{expansions}) in Eq. (\ref{presseqrhoc}). To first order in $\epsilon$ and $x$, we find
\begin{equation}\label{dpdepstrans}
 \frac{d\tilde{p}}{d\epsilon} = \frac{2\Lambda\left(M_{c} -4\pi r_{c}^{3}\Lambda\right)}{r_{c}\left(8\pi r_{c} - 2M_{c}\right)}x - \frac{\left(M_{c} -4\pi r_{c}^{3}\Lambda\right)}{r_{c}\left(8\pi r_{c} - 2M_{c}\right)}\tilde{p}\;.
\end{equation}
Since $\partial\beta/\partial r$ does not appear in Eq. (\ref{dpdepstrans}) we can conclude that a rapid variation from the gCg to the standard Chaplygin gas equation of state at $\rho = \rho_c$ does not change the results obtained so far.

We did not consider here the case of the phantom gCg (i.e. $\rho < \Lambda$) because the standard phantom Chaplygin gas ($\alpha = 1$) is already superluminal by itself [see Eq. (\ref{cs2gcgeq})]. Moreover, note that, in the phantom case, in order that  $c_{s}^2 < 1$ the energy density must satisfy the inequality
\begin{equation}
\alpha^{\frac{1}{\alpha + 1}}\Lambda < \rho < \Lambda\;,
\end{equation}
which implies $\alpha < 1$.

On the other hand, we have shown in Sec. \ref{sec:phant} that the pressure has a singularity at a finite radial distance $r$, where the energy density vanishes. Then we conclude that in the phantom case a superluminal speed of sound cannot be avoided.

\section{Discussion and Conclusions}\label{sec:concl}

In this paper we have investigated solutions of the Tolman-Oppenheimer-Volkoff equations for static and spherically symmetric configurations of the generalized Chaplygin gas, thus generalizing the same analysis carried out in \cite{Gorini:2008zj} for the standard $\alpha = 1$ case. We have found the same geometrical configurations discovered in \cite{Gorini:2008zj}. In the normal (nonphantom) case all solutions, except the de Sitter one, realize a three-dimensional spheroidal geometry because the radial coordinate achieves a maximum value $r_{0}$, dubbed {\it the equator}. After the equator crossing, the same three scenarios studied in \cite{Gorini:2008zj} may take place: a regular spheroid, a truncated spheroid having a scalar curvature singularity at a finite radial distance, and a closed spheroid having a Schwarzschild type singularity at the south pole, namely for $\chi = \pi$. The presence of $\beta$ does not give rise to any new scenario but slightly modifies the pressure and mass profiles in the three scenarios mentioned above, as we have shown by studying their asymptotic behavior.

We have also considered the possibility of a phantom gCg. In this case there is no equator and all solutions have the geometry of a truncated spheroid with the same type of singularity; namely the pressure diverges at a certain finite radius. We point out that the asymptotic behavior of the pressure near the resulting curvature singularity is completely independent of $\beta$. There are only two spherically symmetric static configurations in the presence of the phantom gCg: the first is a truncated spheroid, whereas the second is a wormholelike throat. The peculiar property of the latter is that it does not connect two asymptotically flat space-times. In these cases, the presence of $\beta$ does not affect the qualitative behavior of the pressure. We have also shown that a smooth transition from the generalized to the standard Chaplygin gas near the asymptotic value $p = -\rho = -\Lambda$, required to preserve causality, does not affect the validity of all the previously obtained results.

Finally, let us briefly recapitulate the main similarities and differences between the cases of the standard 
Chaplygin gas and of the gCg. First, the main result of our present investigation is that all geometrical configurations (and only these ones) which we have found in \cite{Gorini:2008zj} for the standard case are also present for the whole range of the gCg parameter $0 < \alpha < \infty$. Second, in spite of the qualitative similarity between the solutions for the gCg and the standard Chaplygin gas, there exists quantitative difference between these cases, especially for the normal (nonphantom) gCg, case 2 (see Sec. \ref{sec:Nonphant}), where the $\alpha$ dependence of the pressure profile and the geometric characteristics acquire a rather involved character. Third, another interesting aspect has been found in the case of the superluminal normal gCg. As we showed in \cite{Gorini:2007ta}, for this case there is a good agreement with observation not only at the level of global cosmological dynamics, but also for the description of perturbations responsible for large scale structure formation. We have shown here that the technique employed in \cite{Gorini:2007ta} to solve the problem of causality in the superluminal fluid could be applied also in the context of the study of static spherically symmetric solutions of the TOV equations.

A natural development of the techniques and results contained in this paper would be the analysis of a starlike object in the standard dynamical gCg universe and of a local spherical collapse of a gCg sphere, the latter process being a kind of generalization of the Lema\^itre-Tolman-Bondi solution \cite{Lemaitre:1933gd, Tolman:1934za, Bondi:1947av}. Though we are fully aware of the difficulties of these problems, we are still hopeful to get some results in this direction in the near future.

\section*{Acknowledgments}

O.F.P. wishes to thank the Institute of Cosmology and Gravitation (ICG), University of Portsmouth, for hospitality during the middle part of this project. A.Y.K. and A.A.S. were partially supported by the RFBR Grant No. 08-02-00923, No. SS-4899.2008.2, and by the Scientific Programme "Astronomy" of the Russian Academy of Sciences.

\bibliographystyle{plain}

\end{document}